\documentclass[11pt]{article}
\usepackage[letterpaper,margin=1in]{geometry}
\usepackage{amsmath,amssymb,graphicx,bm}
\usepackage[hyperfootnotes=false]{hyperref}
\usepackage{booktabs,placeins}
\usepackage{tikz}
\usetikzlibrary{arrows.meta,positioning,calc}
\usepackage{microtype}

\hypersetup{
  colorlinks=true,
  linkcolor=blue,
  citecolor=blue,
  urlcolor=blue,
  pdftitle={EM-Based p-norm-like Constraint RLS Algorithm with Adaptive Regularization for Sparse System Identification},
  pdfauthor={Shuyang Jiang, Kung Yao, Zhenyu Wu}
}

\newenvironment{keywords}{%
  \par\vspace{0.5em}\noindent\textbf{Keywords:}\ }{%
  \par\vspace{0.75em}}

\title{EM-Based $p$-norm-like Constraint RLS Algorithm with Adaptive Regularization for Sparse System Identification}
\author{%
Shuyang Jiang$^{1}$ \qquad Kung Yao$^{1}$ \qquad Zhenyu Wu$^{2,*}$\\[0.6em]
\small $^{1}$Department of Electrical and Computer Engineering,\
\small University of California, Los Angeles, USA\\
\small $^{2}$College of Computer Science and Technology,\
\small National University of Defense Technology, China\\[0.3em]
\small $^{*}$Corresponding author: \texttt{wuzhenyu@nudt.edu.cn}
}
\date{}

\begin{document}
\maketitle
\begingroup
\renewcommand{\thefootnote}{}
\footnotetext{
This work has been submitted to the IEEE for possible publication.
Copyright may be transferred without notice, after which this version
may no longer be accessible.
}
\endgroup
\begin{abstract}
In this paper, the recursive least squares (RLS) algorithm is considered in the sparse system identification setting. The cost function of RLS algorithm is regularized by a $p$-norm-like ($0\leq p\leq 1$) constraint of the estimated system parameters. To minimize the regularized cost function, we transform it into a penalized maximum likelihood (ML) problem that is solved by the expectation-maximization (EM) algorithm, and derive the tap-weight update through a thresholding operator whose sparsity also permits a low-complexity implementation. We further propose an error-driven adaptive regularization scheme operating on two timescales: a fast gate driven by the normalized excess a-priori error relaxes the penalty toward a lower bound, while a slow effective-support tracker sets its stationary target. Numerical simulations demonstrate
the superiority of the new algorithm over the state-of-the-art sparse
RLS algorithms, as well as conventional sparse RLS algorithm.
\end{abstract}
\section{Introduction}
\label{sec:intro}

Adaptive filtering algorithms, such as least mean squares (LMS) and recursive least squares (RLS) algorithms \cite{bib:Haykin}, are widely used in estimation problems where data arrive sequentially. In communication, a wide range of signals of interest is sparse in a certain representation. For example, the multipath wireless channel contains only a few large coefficients among many negligible ones \cite{bib:Bajwa}. However, neither LMS nor RLS exploits the underlying sparseness in order to improve the quality of estimation.

Many LMS-based algorithms have been proposed for sparse system identification. These adaptive algorithms utilize the results from the least absolute shrinkage and selection operator (LASSO) \cite{bib:Tibshirani} and compressive sensing literature \cite{bib:Donoho}, and incorporate the sparsity-promoting norm regularization into the cost function. The first of this kind of algorithms was given in \cite{bib:RaoBD}, where the $\ell_p$ ($0<p\leq 1$) norm of the weight vector is employed as the regularization penalty. Subsequently, $\ell_1$ norm \cite{bib:Chen}, $\ell_0$ norm \cite{bib:Gu} and $p$-norm-like \cite{bib:Wu1} were also utilized to obtain a series of norm constraint LMS algorithms, all of which increase convergence speed and decrease steady-state mean square error (MSE) relative to traditional LMS. This family has since been extended to proportionate schemes unified with sparsity penalties by iterative reweighting \cite{bib:LeeRao}. Robustness to non-Gaussian disturbance is a parallel concern, addressed by soft-root-sign penalties \cite{bib:Patel} and logarithmic hyperbolic-tangent costs \cite{bib:Chikyal25}. A broader account of the underlying nonconvex penalties is given in \cite{bib:Wen}.

RLS based adaptive algorithms are another important class of adaptive algorithms, and a number of RLS-based sparse algorithms have consequently been proposed. Eksioglu and Tanc \cite{bib:Eksioglu2} instead adopt a general sparsity-related convex function as the regularization term; by using results from subgradient calculus, the ensuing convex regularized-RLS (CR-RLS) algorithm retains a form similar to conventional RLS. In \cite{bib:Babadi}, a recursive $\ell_1$-regularized least squares (SPARLS) algorithm is introduced, which is based on an expectation-maximization (EM) type algorithm presented in \cite{bib:Figueirado}. Also based on the EM algorithm, \cite{bib:Liu} addresses in-network distributed estimation of sparse systems using $\ell_1$ and $\ell_0$ norms. In \cite{bib:Zakharov}, a family of low-complexity RLS adaptive filters with different penalties is proposed based on dichotomous coordinate descent (DCD) iterations, an approach later extended to impulsive environments \cite{bib:YuDCD}. Two zero-attracting RLS algorithms \cite{bib:Hong}, ZA-RLS-I and ZA-RLS-II, are derived by employing an adaptively weighted $\ell_2$-norm constraint of the parameter vector. More recent developments pursue robustness and structure rather than new penalties: a unified sparsity-aware robust RLS with jointly optimized parameters handles impulsive noise \cite{bib:YuSRRLS}, and regularization has been revisited through line-search \cite{bib:Stanciu} and tensor-decomposition \cite{bib:Otopeleanu26} formulations. Sparse RLS has likewise been carried into underwater acoustic equalization \cite{bib:Qin,bib:Chen26}. Compared with LMS-based sparse algorithms, RLS-based sparse algorithms have a faster convergence speed and more accurate system estimates.

Two observations motivate the present work. First, the nonconvex regime $0 \leq p < 1$, which is known to approximate the $\ell_0$ measure more tightly \cite{bib:Wen}, remains comparatively unexplored within a recursive least-squares recursion. Second, existing nonconvex treatments reach their solutions through Bayesian machinery \cite{bib:Bai} or unrolled iterations \cite{bib:Monga}, both of which forgo the single-pass recursive structure that makes RLS attractive online. Accordingly, based on the research work on norm constraint LMS and the fact that $p$-norm-like with $0 \leq p \leq 1$ provides an effective mathematical measure of sparsity \cite{bib:Donoho}, we add a $p$-norm-like constraint to the cost function of classic RLS algorithm. Then in a manner alike to the approach in SPARLS \cite{bib:Babadi}, we adopt a maximum likelihood (ML) framework and use the EM algorithm to get the update equation of estimated system parameters. The new algorithm is called EM-$l_{p\textrm{-}\rm{like}}$-RLS algorithm. Moreover, with slight modification, the proposed algorithm can operate with low computational requirement. Inspired by the error-dependent thresholding in HyperLISTA \cite{bib:HyperLISTA}, we further propose a two-timescale error-driven adaptive regularization scheme for the EM-based recursion. A fast gate relaxes regularization in response to normalized excess prediction error, while a slow effective-support tracker adjusts its steady-state target. The numerical simulations show that the proposed algorithm exhibits better estimation performance than the existing sparse RLS algorithms in both stationary and tracking environments.

\section{Background}
\label{sec:background}
In the conventional adaptive filtering setup, the observed output signal $d(n)$ at time $n$ is
\begin{equation}
d(n)=\mathbf{w}^T \mathbf{x}(n) + v(n)
\end{equation}
$\mathbf{w}=[w_0,w_1,...,w_{M-1}]^T$ denotes the sparse impulse response of unknown system and satisfies that $\|\mathbf{w}\|_0 \ll M$, where $\|\mathbf{w}\|_0 = |\{w_i|w_i \neq 0\}|$. ${\mathbf x}(n)=[x(n),x(n-1),...,x(n-M+1)]^T$ is the vector of input signal $x(n)$, and $v(n)$ is the observation noise which is assumed to be i.i.d. Gaussian, i.e., $v(n) \sim \mathcal{N}(0,\sigma^2)$. The estimated tap-weight vector at time $n$ is defined by $\mathbf{\hat{w}}(n)=[\hat{w}_0(n),\hat{w}_1(n),...,\hat{w}_{M-1}(n)]^T.$ Then, the instantaneous error between the desired output and estimated system output is
\begin{equation}
e(n)=d(n)-\mathbf{\hat{w}}^T(n) {\mathbf x}(n),
\end{equation}
and the operation of the filter at time $n$ can therefore be stated as the following optimization problem:
\begin{equation} \label{eq:optimization}
	\min_{\mathbf{\hat{w}}(n)} f(e(1),e(2),...,e(n))
\end{equation}
where $f \geq 0$ is a predefined cost function. A classic choice of sparse RLS algorithm is defined as follows:
\begin{equation} \label{eq:f_CRRLS}
 	f(e(1),\!e(2),\!...,\!e(n))\!=\! \frac{1}{2}\! \sum_{i=1}^{n}\! \lambda^{n-i} e(i)^2 \!+\! \gamma_n g(\mathbf{\hat{w}}(n)),
\end{equation}
where $\lambda$ ($0<\lambda \leq 1$) is an exponential forgetting factor, $g(\cdot)$ is a function that enhances sparsity, and $\gamma_n \geq 0$ is time-varying regularization parameter that governs the trade-off between the estimation error and sparsity of the $\mathbf{\hat{w}}(n)$.

\section{\texorpdfstring{Proposed EM-$l_{p\textrm{-}\rm{like}}$-RLS Algorithm}{Proposed EM-lp-like-RLS Algorithm}}
\label{sec:EMlplikeRLS}
In this section, we will introduce a new approach to derive the update equation of $\mathbf{\hat{w}}(n)$ from $\mathbf{\hat{w}}(n-1)$. In the proposed algorithm, by using the method similar to that of SPARLS \cite{bib:Babadi}, we treat the problem of minimizing the penalized cost function as penalized maximum likelihood problem. Then, we use the EM algorithm to obtain the recursive formula of $\mathbf{\hat{w}}(n)$.

\subsection{\texorpdfstring{The Derivation of EM-$l_{p\textrm{-}\rm{like}}$-RLS Algorithm}{The Derivation of EM-lp-like-RLS Algorithm}}
Let us define
\begin{align}
\mathbf{\Lambda}(n)= &{\rm diag}(\lambda^{n-1},\lambda^{n-2},...,1), \\
\mathbf{d}(n)= &[d(1),d(2),...,d(n)]^T, \\
\mathbf{v}(n)= &[v(1),v(2),...,v(n)]^T, \\
\mathbf{X}(n)= &\left[
\begin{matrix}
\mathbf{x}^T(1)\\
\mathbf{x}^T(2)\\
\vdots \\
\mathbf{x}^T(n)
\end{matrix}
\right].
\end{align}
The penalized cost function \eqref{eq:f_CRRLS} can be rewritten as the following form:
\begin{equation} \label{eq:leastsquare}
f = \frac{1}{2} \| \mathbf{\Lambda}^{1/2}(n) \mathbf{d}(n) - \mathbf{\Lambda}^{1/2}(n) \mathbf{X}(n) \mathbf{\hat{w}}(n)\|_2^2 + \gamma_n \| \mathbf{\hat{w}}(n)\|_{p\textrm{-}\rm{like}},
\end{equation}
where $\mathbf{\Lambda}^{1/2}(n)$ is a diagonal matrix with entries $\Lambda_{ii}^{1/2}(n)=\sqrt{\Lambda_{ii}(n)}$. Here, we set the sparsity-promoting term as the $p$-norm-like of estimated vector $\| \mathbf{\hat{w}}(n) \|_{p\textrm{-}\rm{like}}$. It is well understood that $p$-norm-like is an effective measure of sparsity \cite{bib:Wu1,bib:Wen}, which is defined as
\begin{equation}
\| \mathbf{\hat{w}}(n)\|_{p\textrm{-}\rm{like}} = \sum_{i=0}^{M-1} |\hat{w}_i (n)|^p, \quad 0 \leq p \leq 1.
\end{equation}
Note that $p$-norm-like is a bit different from the classic $\ell_p$ norm. We assume that $|\hat{w}_i (n)|^0 = 0$ when $\hat{w}_i (n)=0$. Then, $p$-norm-like is consistent with $\ell_0$ and $\ell_1$ norm when $p$ is $0$ and $1$.

Here, we adopt a method similar to that presented in \cite{bib:Babadi} and \cite{bib:Figueirado}. The minimization of penalized cost function \eqref{eq:leastsquare} can be identified as the following penalized maximum likelihood problem:
\begin{equation}
\max_{\mathbf{\hat{w}}} \log p(\mathbf{d}(n)|\mathbf{\hat{w}}) - (\gamma_n/\sigma^2) \| \mathbf{\hat{w}}\|_{p\textrm{-}\rm{like}},
\end{equation}
where $p(\mathbf{d}(n)|\mathbf{\hat{w}})=\mathcal{N}(\mathbf{X}(n) \mathbf{\hat{w}}, \sigma^2 \mathbf{\Lambda}^{-1}(n))$ is the likelihood function corresponding to the observation model $\mathbf{d}(n) = \mathbf{X}(n) \mathbf{w} + \bm{\eta}(n)$ with $\bm{\eta}(n) \sim \mathcal{N}(\mathbf{0}, \sigma^2 \mathbf{\Lambda}^{-1}(n))$.

The above ML problem is hard to solve directly. The EM algorithm provides an efficient scheme for solving it. We introduce an auxiliary variable $\mathbf{z}(n) = \mathbf{w} + \alpha \bm{\eta}_1(n)$ with $\bm{\eta}_1(n) \sim \mathcal{N}(\mathbf{0},\mathbf{I})$. $\alpha$ is a parameter to be discussed later. Then the observation model can be rewritten as $\mathbf{d}(n)=\mathbf{X}(n)\mathbf{z}(n)+\mathbf{\Lambda}^{-1/2}(n)\bm{\eta}_2(n)$, with $\bm{\eta}_2(n) \sim \mathcal{N}(\mathbf{0}, \sigma^2\mathbf{I}-\alpha^2\mathbf{\Lambda}^{1/2}(n) \mathbf{X}(n) \mathbf{X}^T(n) \mathbf{\Lambda}^{1/2}(n))$.

We denote $\mathbf{\hat{w}}^{(0)}(n) = \mathbf{\hat{w}}(n-1)$ and $\mathbf{\hat{w}}^{(K)}(n)=\mathbf{\hat{w}}(n)$, where $K$ is the number of iterations. The $l$th ($0\!\leq \!l\! \leq \!K\!-\!1$) iteration of the EM algorithm is as follows:
\begin{equation} \label{eq:EM}
\begin{cases}
{\rm E\textit{-}step:} Q(\mathbf{\hat{w}}, \!\mathbf{\hat{w}}^{(l)}(n)) \!\!\triangleq\!\! \mathbb{E}_z[\log \!p(\mathbf{d}(n), \!\mathbf{z}(n)|\mathbf{\hat{w}})|\mathbf{d}(n), \!\mathbf{\hat{w}}^{(l)}(n)] \\
\qquad \qquad \qquad \qquad = -\frac{1}{2\alpha^2} \|\mathbf{r}^{(l)}(n) - \mathbf{\hat{w}}\|_2^2, \\
\qquad {\rm where\ } \!\mathbf{r}^{(l)}(n)\!=\! \left( \mathbf{I} - \frac{\alpha^2}{\sigma^2} \mathbf{X}^T(n) \mathbf{\Lambda}(n) \mathbf{X}(n)\right) \mathbf{\hat{w}}^{(l)}(n) \\
\qquad \qquad \qquad \qquad \qquad+ \frac{\alpha^2}{\sigma^2} \mathbf{X}^T(n) \mathbf{\Lambda}(n) \mathbf{d}(n), \\
{\rm M\textit{-}step:\ }  \mathbf{\hat{w}}^{(l+1)}(n) \!=\! \arg \max_{\mathbf{\hat{w}}} Q(\mathbf{\hat{w}},\!\mathbf{\hat{w}}^{(l)}(n)) \!-\! \frac{\gamma_n}{\sigma^2} \| \mathbf{\hat{w}}\|_{p\textrm{-}\rm{like}} \\
\qquad \qquad \qquad \qquad = \mathcal{S}(\mathbf{r}^{(l)}(n)),
\end{cases}
\end{equation}
where $\mathcal{S}(\cdot)$ is the element-wise soft thresholding function, which will be derived next. For more detail about the above derivation of the E-step, please refer to \cite{bib:Figueirado}.

\begin{figure}[!t]
	\centering
	\includegraphics[width=0.95\linewidth]{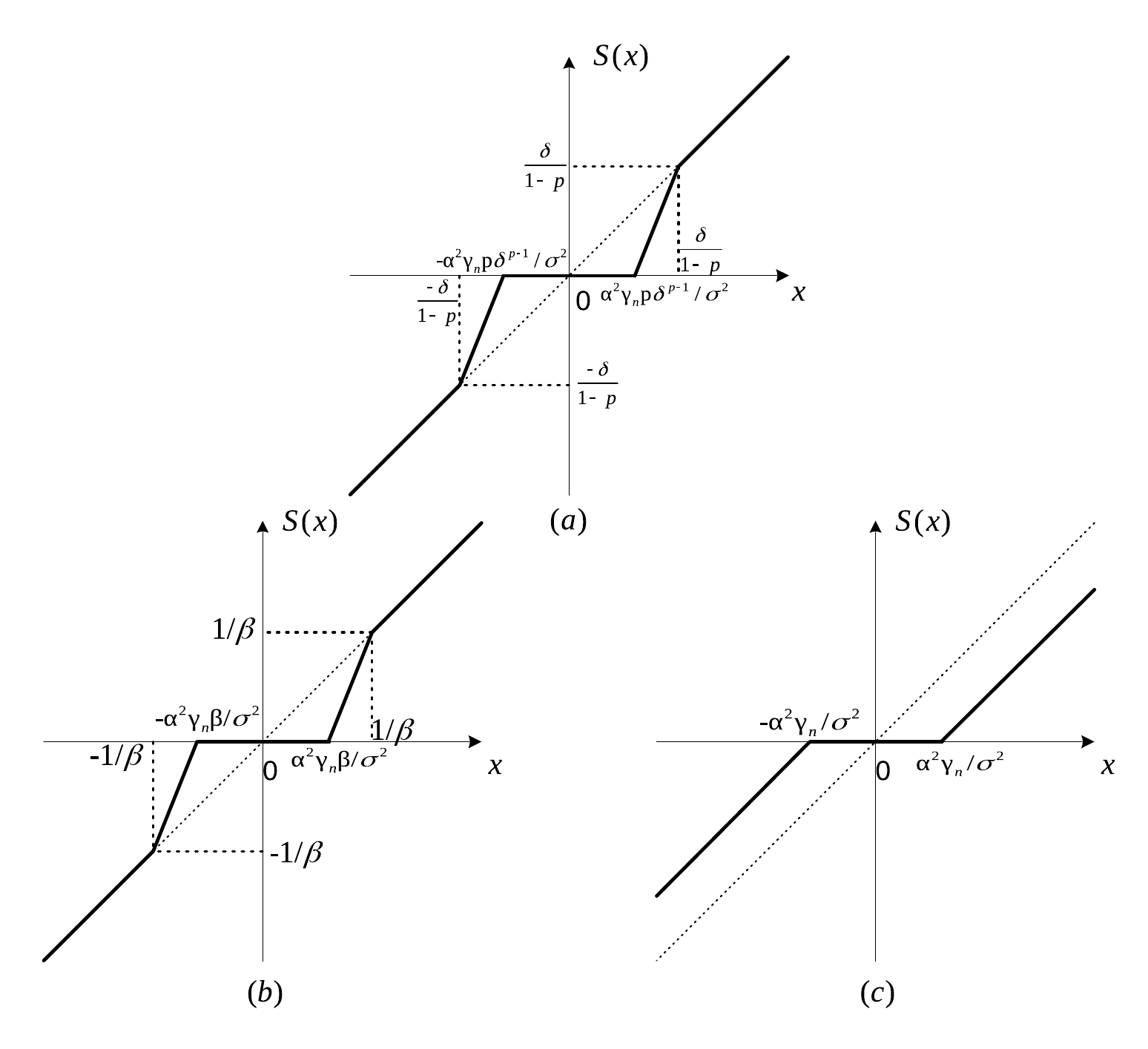}
	\caption{The soft thresholding function of the EM-$l_{p\textrm{-}\rm{like}}$-RLS algorithm, when (a) $0 < p < 1$, (b) $p=0$ and (c) $p=1$.}
	\label{fig:thresholding_func}
\end{figure}

For convenience, all the vector operations are performed element-wise in the derivation of $\mathcal{S}(\cdot)$. To obtain the local maximum of $Q'(\mathbf{\hat{w}}) = Q(\mathbf{\hat{w}},\!\mathbf{\hat{w}}^{(l)}(n)) \!-\! \frac{\gamma_n}{\sigma^2} \| \mathbf{\hat{w}}\|_{p\textrm{-}\rm{like}}$, we consider two cases: $0<p\leq 1$ and $p=0$. When $0<p\leq 1$, we set that $\nabla^s \big( Q'(\mathbf{\hat{w}}) \big) = 0$ and can obtain the element-wise function of $\mathbf{\hat{w}}$ as
\begin{equation} \label{eq:S}
\mathcal{S}'(\mathbf{\hat{w}}) = \mathbf{\hat{w}} + \frac{\alpha^2 \gamma_n}{\sigma^2} \frac{p \ {\rm sgn}(\mathbf{\hat{w}})}{(|\mathbf{\hat{w}}| + \delta \mathbf{1})^{1-p}} = \mathbf{r}^{(l)}(n),
\end{equation}
where $\delta$ is an appropriately selected positive parameter to make $\mathcal{S}'(\cdot)$ invertible, and ${\rm sgn}(\cdot)$ is the sign function. We know that $\mathcal{S}(\cdot) = \mathcal{S}'^{-1}(\cdot)$. Utilizing the approximation that $1/(|x|+\delta)^{1-p}$ is $\delta^{p-1}+(p-1)\delta^{p-2}|x|$ for $x \in (-\frac{\delta}{1-p}, \frac{\delta}{1-p})$ and $0$ otherwise, we can derive the expression of the element-wise $\mathcal{S}(\cdot)$ as shown in Fig.~\ref{fig:thresholding_func} (a) and (c).

When $p=0$, we adopt a popular approximation of $\ell_0$ norm as $\| \mathbf{\hat{w}}\|_0 \approx \sum_{i=0}^{M-1} (1 - e^{-\beta |\hat{w}_i|})$, where $\beta$ is a well-chosen parameter. We set that $\nabla^s \big( Q'(\mathbf{\hat{w}}) \big) = 0$ and then use the approximation that $e^{-\beta |x|}$ is $1\!-\!\beta|x|$ for $|x| \!\leq\! 1/\beta$ and $0$ otherwise, $\mathcal{S}'(\mathbf{\hat{w}})$ defined in \eqref{eq:S} becomes
\begin{equation}
\mathcal{S}'(x) = \begin{cases}
x + \frac{\alpha^2 \gamma_n}{\sigma^2} (\beta {\rm sgn}(x) - \beta^2 x), & |x| \leq 1/\beta \\
x, & {\rm otherwise}
\end{cases}.
\end{equation}
And $\mathcal{S}(\cdot) = \mathcal{S}'^{-1}(\cdot)$ is derived as shown in Fig.~\ref{fig:thresholding_func} (b).

Moreover, we define two new auxiliary variables $\mathbf{B}(n)\!=\! \mathbf{I} \!-\! \frac{\alpha^2}{\sigma^2} \mathbf{X}^T(n) \mathbf{\Lambda}(n) \mathbf{X}(n)$ and $\mathbf{u}(n)\!=\! \frac{\alpha^2}{\sigma^2} \mathbf{X}^T(n) \mathbf{\Lambda}(n) \mathbf{d}(n)$, which have the following rank-one update rules:
\begin{align}
\mathbf{B}(n) &= \lambda \mathbf{B}(n-1) - \frac{\alpha^2}{\sigma^2} \mathbf{x}(n) \mathbf{x}^T(n) + (1-\lambda) \mathbf{I},\label{eq:statistics-b}\\
\mathbf{u}(n) &= \lambda \mathbf{u}(n-1) + \frac{\alpha^2}{\sigma^2} d(n) \mathbf{x}(n).\label{eq:statistics-u}
\end{align}
Note that the soft thresholding function tends to shrink some components of the estimated $\mathbf{\hat{w}}^{(l)}(n)$ to zeros. This fact allows us to restrict the matrix-vector multiplication $\mathbf{B}(n) \mathbf{\hat{w}}^{(l)}(n)$ to the support of $\mathbf{\hat{w}}^{(l)}(n)$ in the E-step of the EM algorithm, and $K$ iterations of the EM algorithm in \eqref{eq:EM} can be implemented in a low computational complexity fashion.

\subsection{Brief Discussion}
\label{sub:briefdiscuss}
Four parameters govern the proposed algorithm. To keep the covariance matrix of $\bm{\eta}_2(n)$ positive semi-definite, $\alpha$ must satisfy $\alpha^2 \!\leq\! \sigma^2/s_1$, where $s_1$ is the largest eigenvalue of the matrix $\mathbf{\Lambda}^{1/2}(n) \mathbf{X}(n)$ $\mathbf{X}^T(n) \mathbf{\Lambda}^{1/2}(n)$; the Gaussian i.i.d. case is analysed further in \cite{bib:Babadi}. We retain $\alpha=\sigma/4$ as the nominal choice. The pair $\beta$ and $\delta$ fixes the range of attraction for small coefficients and must respect $\lim_{p\to0} \delta/(1-p)=\delta=1/\beta$, with $\beta$ taking finite values such as $1$, $5$ or $10$ following \cite{bib:Gu}. The regularization parameter $\gamma_n$ controls the tradeoff between output sparseness and estimation error; the selection of $\gamma_n$ will be discussed in detail in Section \ref{sec:adaptive}. A theoretical selection rule can be referred in \cite{bib:Liu}. Finally, $K = 1$ EM iteration suffices for convergence, as confirmed by our simulations and by \cite{bib:Babadi, bib:Liu}, and it also reduces the computational complexity to the minimum.

\subsection{Two-Timescale Error-Driven Adaptive Regularization}
\label{sec:adaptive}
\begin{figure}[!t]
  \centering
  \includegraphics[width=0.72\linewidth]{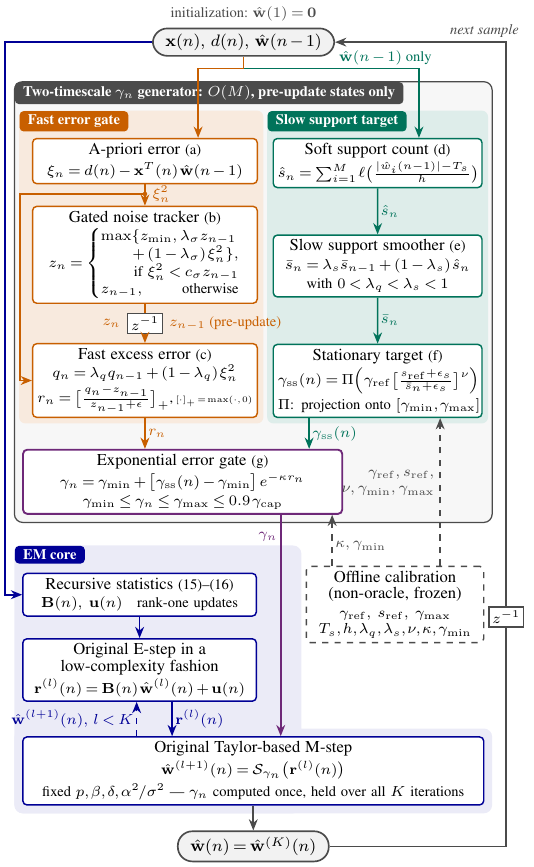}
  \caption{The complete procedure for the EM-$l_{p\textrm{-}\rm{like}}$-RLS algorithm.}
  \label{fig:architecture}
\end{figure}

Inspired by error-dependent thresholds in HyperLISTA \cite{bib:HyperLISTA} and the works of \cite{bib:Eksioglu2, bib:Babadi, bib:Jin} that treat weight selection as a separate design problem, we
propose a two-timescale error-driven adaptive regularization scheme, which is summarized in the upper part of Fig.~\ref{fig:architecture}.

The fast error gate branch is designed to detect the prediction error. It first computes a-priori error $\xi_n$ (defined in Fig.~\ref{fig:architecture}(a)). The fast error-energy $q_n$ (defined in Fig.~\ref{fig:architecture}(c)) smooths instantaneous a-priori error while retaining the previous memory. Since the noise variance is unknown, we estimate it through Fig.~\ref{fig:architecture}(b). Then we compare the error-energy $q_n$ with the previous noise variance estimate $z_{n-1}$ (defined in Fig.~\ref{fig:architecture}(b)), and get the normalized excess-error $r_n$ (defined in Fig.~\ref{fig:architecture}(c)), which is a good measure of the prediction error. $r_n$ detects excess residual energy: coefficient changes, noise changes, outliers and persistent bias can all increase it.

The slow support target branch determines the stationary regularization level from the estimated sparsity structure. The soft support count $\hat{s}_n$ (defined in Fig.~\ref{fig:architecture}(d)) replaces a hard threshold with a logistic function of $|\hat{w}_i(n-1)|$. Here, a fixed threshold $T_s>0$, smoothing width $h>0$ and logistic function $\ell(a)=(1+e^{-a})^{-1}$. The smoother $\bar{s}_n$ (defined in Fig.~\ref{fig:architecture}(e)) with $\lambda_s>\lambda_q$ gives the support estimate a longer memory than the error branch. The stationary regularization target $\gamma_{\rm ss}(n)$ (defined in Fig.~\ref{fig:architecture}(f)) raises the regularization when the estimate is sparser than the reference and lowers it when more coefficients become active, with projection onto $[\gamma_{\min},\gamma_{\max}]$ keeping the target bounded.

The exponential error gate combines the two branches through Fig.~\ref{fig:architecture}(g). So the support branch sets the stationary regularization target and the error branch controls how much of it is applied.When $r_n=0$, the regularization equals the stationary target, and as the normalized excess-error grows, it decays smoothly toward $\gamma_{\min}$. Relaxing rather than strengthening the penalty during large excess-error avoids the dead zones of the thresholding operators suppressing newly active coefficients, which is the main risk after an abrupt system change; once the error subsides, the weight returns toward its stationary target and the sparsity pressure is restored.

Offline calibration supplies frozen reference quantities rather than oracle information. Here $\epsilon>0$, $\epsilon_s>0$, $\nu,\kappa>0$, and $0\leq\gamma_{\min}\leq\gamma_{\rm ref}\leq\gamma_{\max}<\infty$. $z_0$ is initialized from a separate noise-only or justified stationary calibration segment. We select $\gamma_{\mathrm{ref}}$ by minimizing the calibration-window MSE over a predefined grid on independent calibration data, and obtain $s_{\mathrm{ref}}$ as the median of the smoothed soft-support counts after burn-in from separate stationary reference runs using this fixed $\gamma_{\mathrm{ref}}$. Initialize $q_0=z_0$, $\bar s_0=s_{\rm ref}$ and $\gamma_0=\gamma_{\rm ref}$ immediately before the first processed adaptive sample.

Finally, we give out the complete procedure for the EM-$l_{p\textrm{-}\rm{like}}$-RLS algorithm in Figure~\ref{fig:architecture}.

\FloatBarrier
\section{Experimental Evaluation}
\label{sec:simulation}
Numerical experiments compare EM-$l_{p\textrm{-}\rm{like}}$-RLS with RLS, CR-RLS-$\ell_1$ and CR-RLS-$\ell_0$ \cite{bib:Eksioglu2}, ZA-RLS-II \cite{bib:Hong}, $\ell_1$-PRLS \cite{qin2021proportionate}, and S-RRLS \cite{bib:YuSRRLS}.
Each experiment uses length-$100$ systems with uniformly sampled supports and unnormalized $\mathcal N(0,1)$ nonzero coefficients, held fixed across trials.
Independent Gaussian inputs and observation noise have variances $0.01$ and $0.005$, respectively; the noise variance is unknown to the algorithms.
All methods use $\lambda=0.999$, and coefficient $\mathrm{MSE}(n)=\mathbb E\|\widehat{\mathbf w}(n)-\mathbf w(n)\|_2^2$ is averaged over $200$ paired trials.
The proposed EM variants use $\beta=5$, $\delta=0.2$, $K=1$, $\epsilon=10^{-8}$, and $\epsilon_s=10^{-6}$, with online noise-estimation parameters $\lambda_\sigma=0.999$, $c_\sigma=9$, and $z_{\min}=10^{-10}$.
The initial noise estimate is calibrated from stationary residuals; the remaining controller parameters are selected by coefficient MSE on independent calibration data and frozen before testing.
For the competing algorithms, all the relevant parameters are fine-tuned to get the best performance on independent calibration data within the prescribed search grids.

\begin{figure}[!t]
	\centering
	\includegraphics[width=\linewidth]{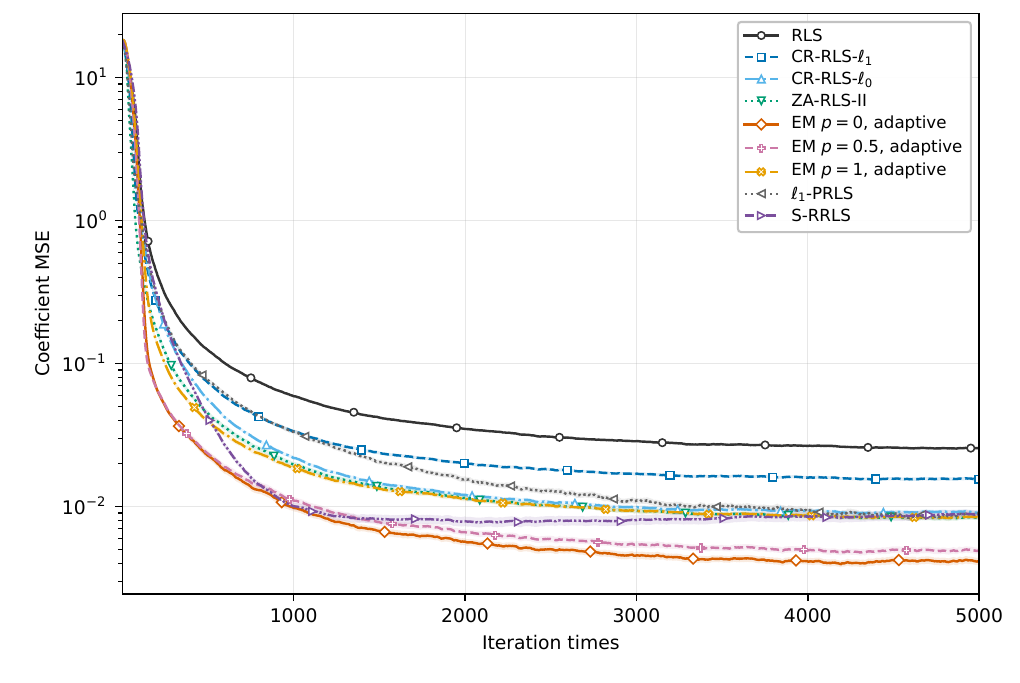}
	\caption{Stationary coefficient-MSE learning curves for nine configurations on one ten-taps sparse system. Lines average $200$ paired trials; All three EM curves use adaptive regularization with noise variance estimated online.}
	\label{fig:learning_curve_sigmav005}
\end{figure}

The first experiment identifies one ten-taps sparse system over $5000$ observations, using $p=0,0.5,1$ and a constant dimensionless EM step of $1/16$.
Parameters are calibrated using mean MSE over the final $1000$ observations of independent stationary streams.
Figure~\ref{fig:learning_curve_sigmav005} compares nine learning curves: adaptive EM with $p=0$ and $p=0.5$ achieves the lowest steady-state MSEs, while $p=1$ performs similarly to ZA-RLS-II.
Thus, the nonconvex variants provide favorable stationary estimates.

\begin{figure}[!t]
\centering
\includegraphics[width=\columnwidth]{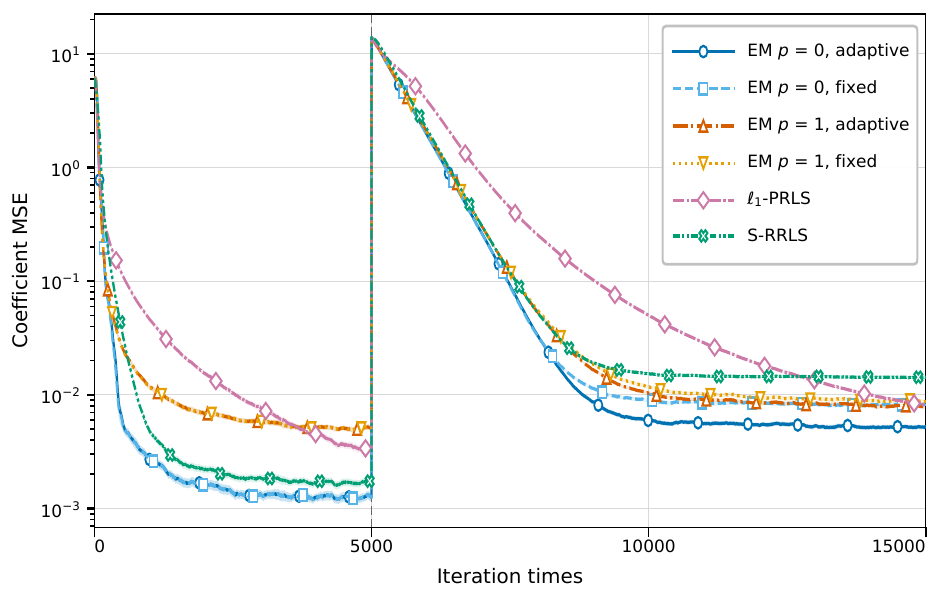}
\caption{Mean coefficient MSE for six configurations in the tracking environment; the vertical line marks the disjoint five-to-ten-tap change at observation $5001$.}
\label{fig:j510}
\end{figure}

The second experiment switches between disjoint five- and ten-tap systems at observation $5001$, without resetting the algorithms or supplying the true support or change time.
Adaptive and fixed EM with $p=0,1$ are compared with $\ell_1$-PRLS and S-RRLS.
Figure~\ref{fig:j510} shows the abrupt error increase and subsequent recovery: adaptive $p=0$ reaches the lowest final MSE, $\ell_1$-PRLS recovers more slowly, and S-RRLS retains higher error.
Both adaptive variants reduce whole-post MSE relative to their fixed counterparts, with clearer gains late in recovery.

\section{Conclusion}
\label{sec:conclusion}
In this paper, we introduce a new sparse RLS algorithm for sparse system estimation using the $p$-norm-like function as the sparsity penalty. The update equations are derived using the EM algorithm within a penalized maximum likelihood framework. We further develop a two-timescale error-driven adaptive regularization scheme to adjust the penalty strength online. A fast gate responds to normalized excess prediction error, while a slow effective-support tracker determines the steady-state regularization target. Simulation results show that the adaptive nonconvex variants achieve favorable estimation performance relative to the compared sparse RLS algorithms.

\bibliographystyle{IEEEbib}
\bibliography{refs,experiment-refs}

\end{document}